\documentclass[aps,pra,groupedaddress,twocolumn]{revtex4-1}
\usepackage{color}
\def\myscalea{1}

\usepackage{graphicx}

\usepackage{amsmath}

\begin{document}

\title{Exact bidirectional X-wave solutions in fiber Bragg gratings}

\author{Nikolaos K. Efremidis}
\email{Corresponding author: nefrem@uoc.gr}
\affiliation{Department of Mathematics and Applied Mathematics, University of Crete, 70013 Heraklion, Crete, Greece}
\author{Nicholas S. Nye}
\author{Demetrios N. Christodoulides}
\affiliation{CREOL/College of Optics, University of Central Florida, Orlando, Florida 32816}

\date{\today}

\begin{abstract}
We find exact solutions describing bidirectional pulses propagating in fiber Bragg gratings. They are derived by solving the coupled-mode theory equations and are expressed in terms of products of modified Bessel functions with algebraic functions. Depending on the values of the two free parameters the general bidirectional X-wave solution can also take the form of a unidirectional pulse. We analyze the symmetries and the asymptotic properties of the solutions and also discuss about additional waveforms that are obtained by interference of more than one solutions. Depending on their parameters such pulses can create a sharp focus with high contrast.
\end{abstract}


\maketitle

\section{Introduction}

Pulse propagation in fiber Bragg gratings is associated with phenomena that arise from the presence of Bloch waves and are utilized in a variety of applications~\cite{kashyap1999fiber,agrawal-applications}. 
In the case of relatively weak index modulations and for optical frequencies close the Bragg frequency coupled-mode theory can be successively applied~\cite{kogel-jap1972}.
The characteristic properties of their spectrum are utilized in a variety of applications. In particular, they are widely used as filters in optical communication systems, including distributed feedback gratings~\cite{haus-ieeejqe1976,utaka-el1984}, Fabry--Perot interferometers~\cite{legoub-josaa1995} and Michelson interferometers~\cite{hill-el1987} among others. In addition a variety of fiber lasers have been developed by utilizing fiber gratings. In this case, the free space mirrors are replaced with a pair of fiber Bragg gratings thus eliminating the need for realignment. In the nonlinear regime the pulse dynamics can lead to soliton formation~\cite{chris-prl1989,aceve-pla1989} as well as to pulse compression~\cite{winfu-apl1985}. 

Exact X-wave solutions were first predicted in the area of Ultrasonics~\cite{lu-tuf1992a,lu-tuf1992b}. Linear~\cite{saari-prl1997,porra-ol2003} as well as nonlinear X-waves~\cite{conti-prl2003x} have been predicted and observed in optics. Exact X-wave solutions of the Bessel type have also been found~\cite{chris-ol2004}. In the nonlinear regime exact X-wave solutions are supported in the presence of an index potential~\cite{efrem-pla2009}. Dark nonlinear X-waves have recently been predicted~\cite{baron-ol2016}.

In this article we find exact solutions describing bidirectional pulses propagating in fiber Bragg gratings. We rely on analytically solving the underlying system of coupled-mode theory equations. Our solution takes the form of a product of a modified Bessel function with an algebraic term for both the forward and the backward components.
Depending on the values of its two free parameters, the general bidirectional X-wave solution can also take the form of a forward or a backward unidirectional pulse. 
We study the symmetries and the asymptotic properties of these solutions as well as their interference. The pulse carries infinite energy since at its tails the amplitude decay with time as $1/\sqrt{|t|}$. However finite energy implementations are possible by truncation or apodization by multiplying the solution with a slowly varying finite energy pulse. It is interesting to point out that depending of the parameters such pulses can create a focus inside the grating.

\section{Coupled-mode theory equations}

We assume that the dominant term in the refractive index distribution 
$\overline n+\delta n\cos(2\pi  z/\Lambda)+n_2I$ is the average value of the index $\overline n$, meaning that we have weak index modulations $\overline n\gg\delta n$ and relatively weak nonlinearities $\overline n\gg n_2I$, where $n_2$ is the Kerr nonlinear index coefficient, and $I$ is the intensity of light. In addition, we assume low pulse intensities so that nonlinear index variations can be ignored.
Then for optical frequencies close to the Bragg frequency, the dynamics of the forward $A_f$ and the backward $A_b$ components inside a fiber grating are described by the following coupled-mode theory differential equations~\cite{kogel-jap1972}
\begin{align}
i(\partial_z +\beta_1\partial_t ) A_f
+\delta A_f+\kappa A_b & = 0 
\label{eq:bragg1}
\\
i(-\partial_z+\beta_1\partial_t) A_b
+
\delta A_b+\kappa A_f & = 0,
\label{eq:bragg2}
\end{align}
where $\kappa>0$ is the coupling coefficient between the forward and the backward waves, $\delta$ is a measure of the detuning from the Bragg frequency $\delta=\beta_1(\omega_0-\omega_B)$, and $\beta_1=1/v_g=\overline n/c$ is inversely proportional to the group velocity inside the fiber without the grating. 
Assuming that both $z$ and $t$ lie on the real line, Eqs.~(\ref{eq:bragg1})-(\ref{eq:bragg2}) support plane-wave solutions of the form $e^{ikz-i(\omega-\omega_0) t}$ that satisfy the following dispersion relation
\begin{equation}
k = \pm\sqrt{D^2-\kappa^2}, 
\end{equation}
with $D=\beta_1(\omega-\omega_B)$.
We note that for 
$-\kappa<D<\kappa$ a gap in the spectrum exists where reflection is dominant in the case of a finite grating.

\section{Exact solutions and their properties}
Using standard methods of partial differential equations it can be shown that the system of Eqs.~(\ref{eq:bragg1})-(\ref{eq:bragg2}) supports the following composite exact solution
\begin{equation}
\Psi_m = (A_f,A_b) = (W_f,W_b)e^{i\delta v_g t},
\label{eq:sol:01}
\end{equation}
where, $W_f$, $W_b$ are the amplitudes of the two components without the additional phase term that are related to $W_m$ via
\begin{equation}
V_m=(W_f,W_b)=(W_m,W_{m-1}),
\label{eq:sol:02}
\end{equation}
with $V_m$ being the composite (vector) form, 
\begin{equation}
W_m = 
(-1)^mUK_m\left[\kappa\sqrt{z^2-\left(\frac{t-it_0}{\beta_1}\right)^2}\right]
\left[
\frac{t-it_0+\beta_1z}{t-it_0-\beta_1z}
\right]^{\frac m2},
\label{eq:sol:03}
\end{equation}
$K_m$ is the modified Bessel function of order $m$, and $U$ is the amplitude.
The solutions depend on two free parameters: the dimensionless variable $m$ that describes the order of the solution, and the real variable $t_0$ with time dimensions. These solutions are spatiotemporally localized, and thus the field amplitude goes to zero as either $z\rightarrow\pm\infty$ or $t\rightarrow\pm\infty$.

By introducing normalized and dimensionless coordinates $\tau$, $\zeta$ and the normalized and dimensionless parameters $\mu$, $\Delta$, $B$ (that are directly related to $t_0$, $\delta$, $U$, respectively)
\begin{equation}
\tau = \frac{\kappa t}{\beta_1},\ 
\zeta=\kappa z,\ 
\mu = \frac{\kappa t_0}{\beta_1},\ 
\Delta = \frac\delta\kappa,\ 
B = \frac U{\sqrt{P_0}},
\end{equation}
where $P_0$ is a reference value for the power, the above exact solution can be written as
\begin{equation}
\Psi_m = V_m e^{i\Delta\tau} = (W_m,W_{m-1}) e^{i\Delta\tau},
\label{eq:sol:norm:01}
\end{equation}
where
\begin{equation}
W_m = (-1)^mB
K_m\left[\sqrt{\zeta^2-\left(\tau-i\mu\right)^2}\right]
F_m(\tau,\zeta,\mu),
\label{eq:sol:norm:02}
\end{equation}
and
\begin{equation}
F_m(\tau,\zeta,\mu) = 
\left[\frac{\tau-i\mu+\zeta}{\tau-i\mu-\zeta}\right]^{\frac m2}.
\label{eq:sol:norm:03}
\end{equation}
The advantage of the normalized Eqs.~(\ref{eq:sol:norm:01})-(\ref{eq:sol:norm:03}) is that the additional parameters $\kappa$ and $\beta_1$ are absorbed in the transformations. Note that $\beta_1=1/v_g$ determines the maximum value of the group velocity inside the grating. On the other hand, changing $\kappa$ modifies the argument of the Bessel function which, in turn, changes the width of the solutions as well as contrast at the focus. Thus our solution depends on two free parameters $\mu$ and $m$.

Assuming that the length of the Bragg grating is $2z_0$ (extending between $-z_0$ and $z_0$) pulses with a profile that is designed according to our solutions should excite the two edges of the grating. Specifically, the forward propagating part should be excited from $z=-z_0$ whereas the backward propagating part should be excited from $z=z_0$.

\begin{figure}
\centerline{\includegraphics[width=\myscalea\columnwidth]{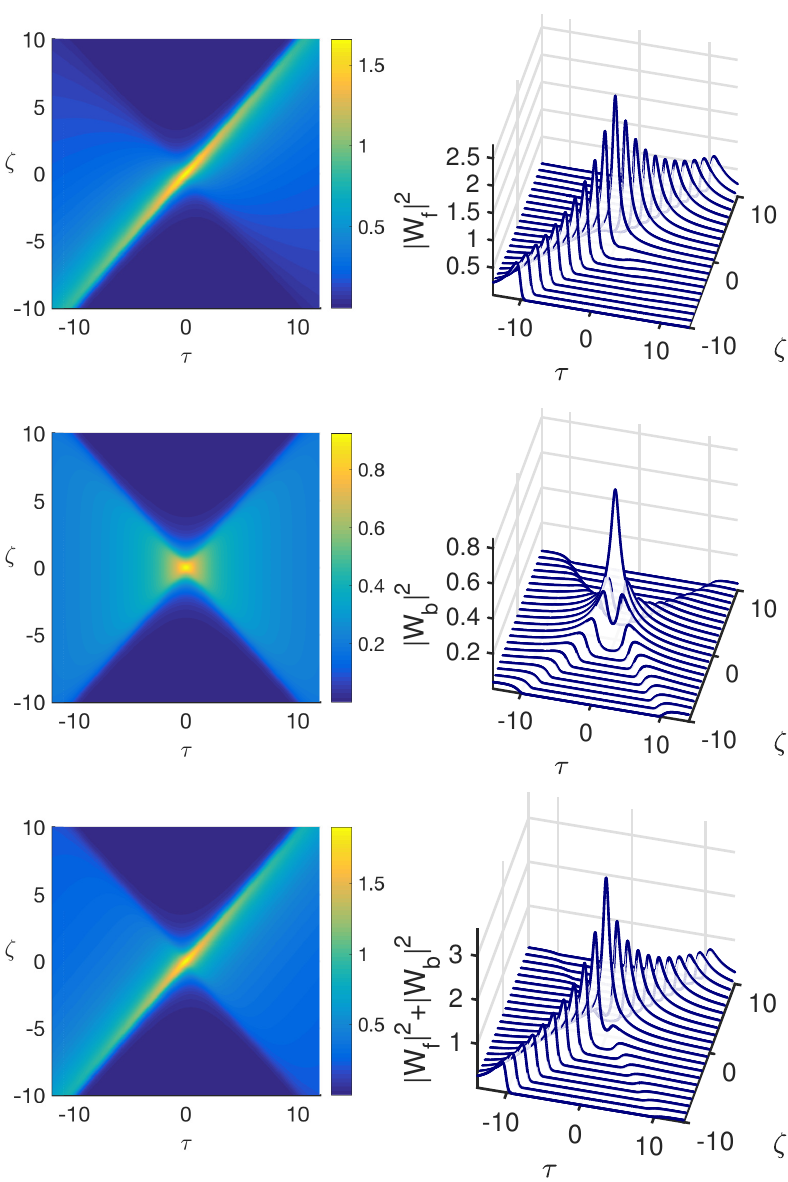}}
\caption{Contour plot of the field amplitude dynamics (left column) and the power dynamics (right column) for $m=1$, $\mu=0.5$, $B=1$. In the three rows we depict the forward, the backward, and the total pulse components, respectively. \label{fig:1}}
\end{figure}
\begin{figure*}
\centerline{\includegraphics[width=\textwidth]{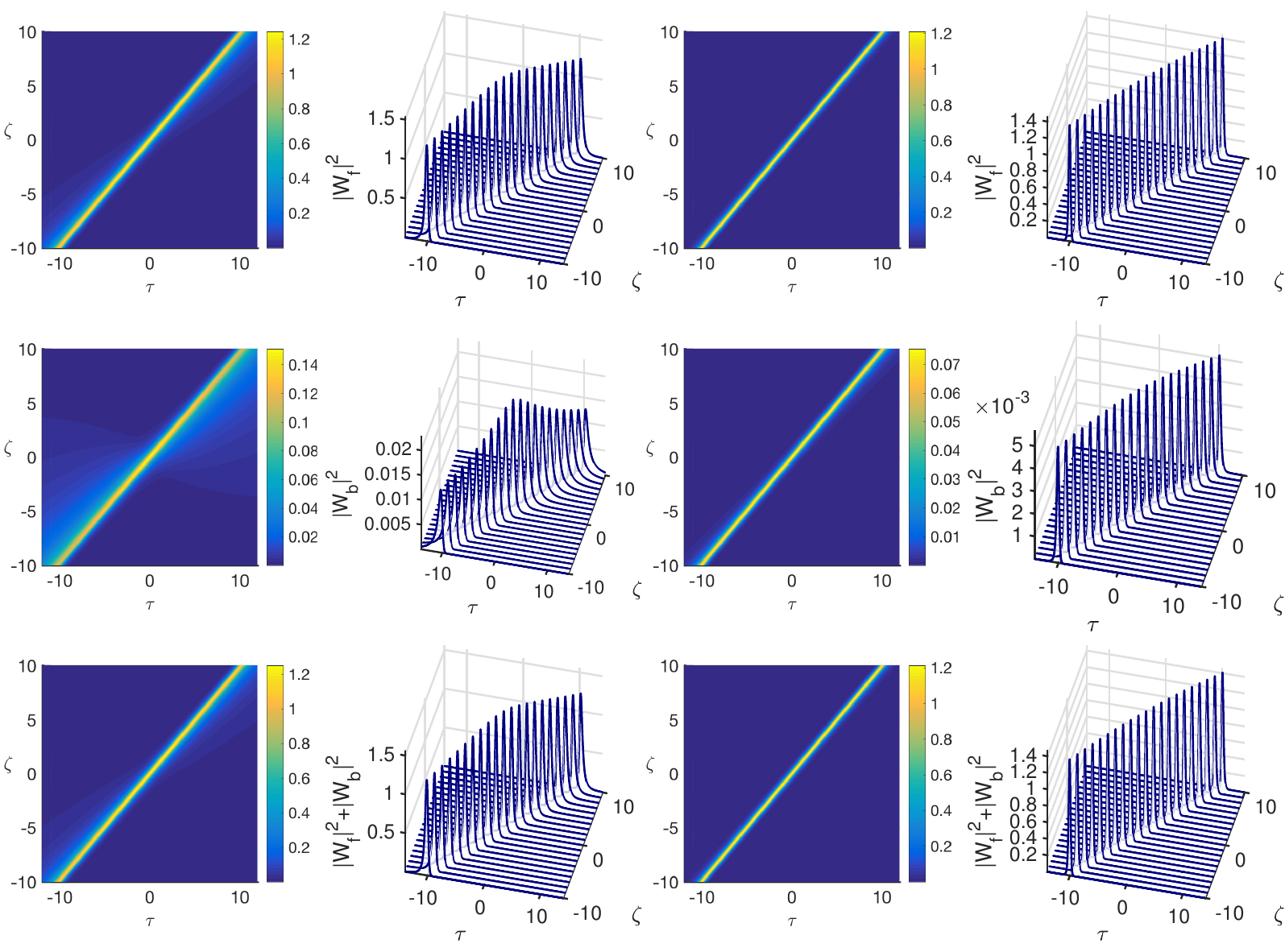}}
\caption{Contour plot of the field amplitude dynamics (1st and 3rd columns) and the power dynamics (2nd and 4th column) for $m=3$, $\mu=0.5$, $B=0.02$ (1st and 2nd column) and $m=5$, $\mu=0.5$, $B=0.0001$ (3rd and 4th columns). In the three rows we depict the forward, the backward, and the total pulse components, respectively. \label{fig:2}}
\end{figure*}

\subsection{Symmetries}
We are now going to explore the symmetries of the system. The first property is derived from the system of Eqs.~(\ref{eq:bragg1})-(\ref{eq:bragg2}). Specifically, if $A(\tau,\zeta)$ is any solution of Eqs.~(\ref{eq:bragg1})-(\ref{eq:bragg2}) then $A^*(-\tau,-\zeta)=A(\tau,\zeta)$. In our case this directly translates to 
\begin{equation}
V_m(\tau,\zeta,\mu) = V_m^*(-\tau,-\zeta,\mu).
\label{eq:prop01}
\end{equation}
It is obvious that by making the substitution $\mu\rightarrow-\mu$ is equivalent to complex conjugation, i.e., 
\begin{equation}
V_m(\tau,\zeta,-\mu) = V_m^*(\tau,\zeta,\mu).
\end{equation}
Thus the amplitude of the solution is independent on the sign of $\mu$ and $|V_m(\tau,\zeta,-\mu)| = |V_m(\tau,\zeta,\mu)|$. For this reason we can restrict to the case of positive $\mu$. Since $K_m(z)=K_{-m}(z)$, the effect of the substitution $m\rightarrow-m$ affects only the algebraic term of the solution. Specifically, the relation $W_{-m}(\tau,\zeta,\mu) = W_m(\tau,-\zeta,m)$ translates to 
\begin{equation}
V_{-m}(\zeta) = \sigma_x V_{m+1}(-\zeta)
\end{equation}
where $\sigma_x$ is a Pauli matrix and shows that by reversing the sign of $m$ is equivalent to space inversion (or parity). Thus, we can restrict to the case $m\ge0$. Noting that the substitution $\tau\rightarrow-\tau$ leads to
\begin{equation}
K_m[(\zeta^2-\left(\tau+i\mu\right)^2)^{1/2}]=
K_m^*[(\zeta^2-\left(\tau-i\mu\right)^2)^{1/2}]
\end{equation}
we conclude that the amplitude of the solution associated with the modified Bessel function is even with respect to $\tau$. Thus, any asymmetries in the amplitude of the temporal profile (see Figs.~\ref{fig:1}-\ref{fig:2}) are related to the presence of the algebraic term $F_m$. 

\subsection{Examples}
In Figs.~\ref{fig:1}-\ref{fig:2} we see typical examples of the exact solutions. Specifically, in Fig.~\ref{fig:1} $m=1$ and $|W_f|$ is highly asymmetric with the forward propagating part being significantly stronger as compared to the backward propagating part. On the other hand, $W_b=W_0$ the algebraic term is constant (unity) and the solution is even with respect to $\tau$ ($|W_b(\tau)|=|W_b(-\tau)|$). As a result the spatiotemporal dynamics of $W_f$ take the form of a symmetric (even) X-wave, consisting of two pulses that carry the same amount of energy, one propagating to the right direction and one propagating to the left direction. As a result the total power $|W_f|^2+|W_b|^2$ takes the form of an asymmetric X-wave with the forward propagating component carrying more energy as compared to the backward propagating component. Focusing on $W_f$ we see that its profile close to the peak power is highly asymmetric taking the form of a shock wave or a caustic. Specifically, for negative $\zeta$ the solution is very steep in the right edge of the shock decaying abruptly to zero as $\tau$ increases. On the other hand the solution decays slowly to zero as $\tau$ decreases. The direction of the shock is inverted for $\zeta>0$ as expected from Eq.~(\ref{eq:prop01}). $W_b$ also exhibits similar behavior along both of its high power peaks. Note that the maximum power increases up to $\zeta=0$ ($\zeta=0$, $\tau=0$ is the focus) and then decreases as $\zeta$ starts to take positive values. We see that close to $\zeta=0$, $\tau=0$ the pulse focuses and its power increases. This focusing effect can be significantly enhanced by increasing the value of $\mu$ leading to a very sharp focus with high contrast.

Two additional examples for $m=3$ and $m=5$ are shown in Fig.~\ref{fig:2}. We can clearly see in this case that both pulse components co-propagate to the right, with $W_f$ being the dominant in terms of its maximum power. In comparison with Fig.~\ref{fig:1} the width of the pulse is narrower decaying much faster to zero. In addition, the maximum power does not change as significantly with $\zeta$ as in the previous example. Note that as $m$ increases the solutions become more unidirectional, the relative amplitude of $W_f$ decreases, and the solutions become more localized.

\subsection{Asymptotics calculations}
The properties of these pulses can be analyzed by carrying out asymptotic calculations. We focus in the case where the absolute value of the argument of the Bessel function is large enough. Note that for constant $\zeta$ we have $|\zeta^2-\left(\tau-i\mu\right)^2|^{1/2}\ge\sqrt{2\mu|\zeta|}$. Thus for $|\zeta|\gg1/(2\mu)$ the following expression
\begin{equation}
W_m = \sqrt{\frac\pi2}
\frac{e^{-(\zeta^2-(\tau-i\mu)^2)^{1/2}}}
{(\zeta^2-\left(\tau-i\mu\right)^2)^{1/4}}
\left[
\frac{\mu+i(\tau+\zeta)}{\mu+i(\tau-\zeta)}
\right]^{\frac m2}
\label{eq:asym}
\end{equation}
is highly accurate for all $\tau$. Note that in this limit asymptotically the Bessel function becomes independent of its order $m$ and all the information about the order of the solution is carried out by the algebraic term. From the absolute value of Eq.~(\ref{eq:asym}) we obtain
\begin{multline}
|W_m|= 
\sqrt{\frac\pi2}
\frac{e^{-((\zeta^2+\mu^2-\tau^2)^2+(2\mu\tau)^2)^{1/4}
\cos\frac12\arg(\zeta^2+(\mu+i\tau)^2)}
}
{((\zeta^2+\mu^2-\tau^2)^2+(2\mu\tau)^2)^{1/8}}
\\
\times\left[
\frac{\mu^2+(\tau+\zeta)^2}{\mu^2+(\tau-\zeta)^2}
\right]^{\frac m4}
\label{eq:asampl}
\end{multline}

\begin{figure}
\centerline{\includegraphics[width=\myscalea\columnwidth]{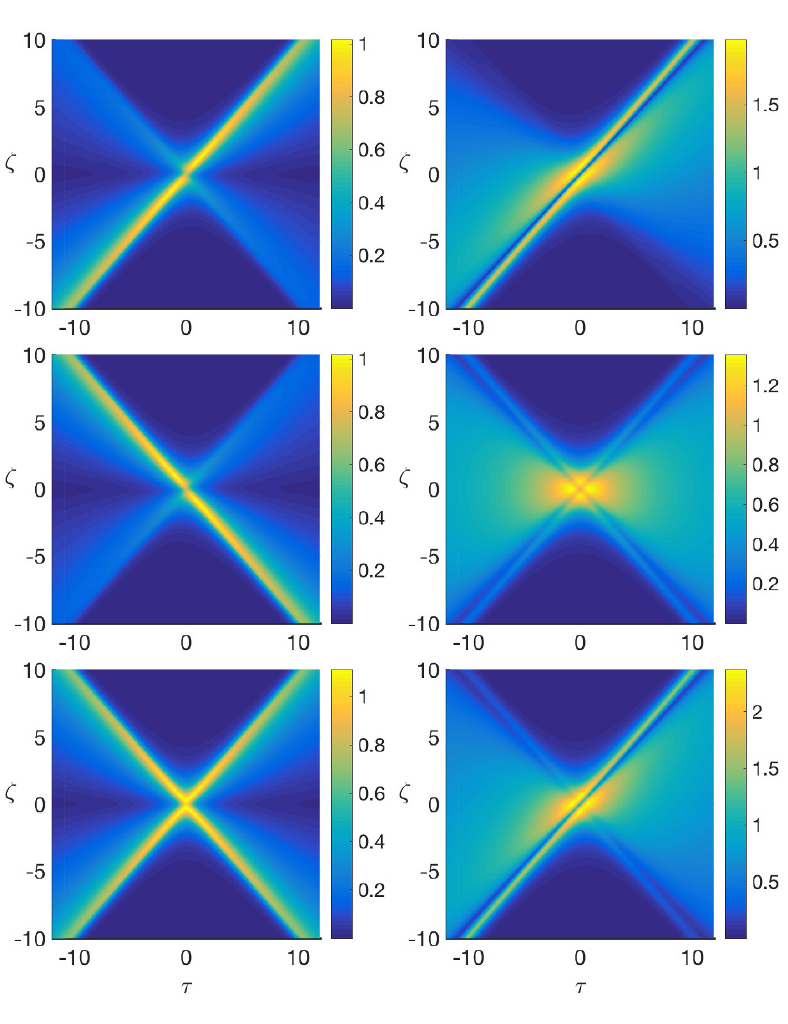}}
\caption{Interference of two pulses $A_m$ with $m=1$, $\mu=0.5$ $B=1$ and $m=0$, $\mu=0.5$, $B=-1$ (left column) and $m=1$, $\mu=0.8$ $B=10$ and $m=1$, $\mu=0.5$, $B=-5$ (right column).  In the three rows we depict $|W_f|$, $|W_b|$, and $\sqrt{|W_f|^2+|W_b|^2}$, respectively.\label{fig:3}}
\end{figure}

By taking the limit $|\tau|\gg|\zeta|$ we can find the asymptotic behavior of the solution at its tails. Specifically, keeping only first order terms we find that 
\begin{equation}
W_m\approx \sqrt{\frac\pi2}\frac{e^{i\tau-\mu}}{(i\tau)^{1/2}}
\label{eq:tails}
\end{equation}
Since the solution decays as $1/\sqrt{|\tau|}$ the total energy carried by the pulse is infinite. However, as in the case of diffraction-free waves (of the Airy or the Bessel type) finite energy realizations that maintain the properties of the solutions can be derived. This can be done either by multiplying the initial condition with a broad pulse with finite energy (for example with a Gaussian or a flat-top super-Gaussian) or by truncating the initial condition (setting its amplitude to zero) for $|\tau|>|\tau_0|$. 

The free parameter $\mu$ strongly affects the maximum values of the amplitude of the solution. Specifically, we see that as $\mu$ goes to zero, both the Bessel and the algebraic part of the solution become singular with the amplitude going to infinity along the trajectory $\tau=\zeta$. In addition, the pulsewidth gets reduced. On the other hand, by increasing $\mu$ the pulsewidth increases and the peak power reduces. 

The second free parameter of our solution is the order $m$. As we showed for  $|\zeta|\gg1/(2\mu)$ all the information about the order of the solutions is carried out by the algebraic term $F_m$. Thus, in order to analyze the effect of the parameter $m$ we will focus on finding the extrema of $|F_m|$. Specifically, $|F_m|$ has two extrema
\begin{equation}
\tau_\pm=\pm\sqrt{\zeta^2+\mu^2} 
\label{eq:taupm}
\end{equation}
which, via comparison with the numerical results coincide with the trajectories of the forward and backward propagating pulses. When $m>0$ and $\tau>0$ the trajectory $\tau_+(\zeta)$ of Eq.~(\ref{eq:taupm}) is a maximum of $|F_m|$ whereas $\tau_-(\zeta)$ is a minimum. By changing the sign of either $m$ or $\tau$ results to interchanging the locations of the minimum and the maximum. Close to the maximum the algebraic term becomes
\begin{equation}
|F_m| \approx
\left(
\frac{4\tau\zeta}
{\mu^2+(\tau-\zeta)^2}
\right)^{m/4}.
\end{equation}
Close to the minima in the above formula the numerator and the denominator are interchanged. We see that as $m$ increases the amplitude close the maximum of $|F_m|$ increases rapidly, whereas at the minimum the amplitude of $|F_m|$ goes to zero. Thus by increasing $m$ the solution becomes more unidirectional and, in addition, more temporally localized. The effect of increasing $m$ while keeping a constant value of $\mu$, as described above, becomes apparent by comparing Fig.~\ref{fig:1} with Fig.~\ref{fig:2}. Furthermore, since the amplitude at the tails, as given by Eq.~(\ref{eq:tails}) does not depend on $m$ the contrast between the maximum pulse power and the power at the tails increases with $m$. 

\subsection{Interference of solutions}

Eqs.~(\ref{eq:sol:norm:01})-(\ref{eq:sol:norm:03}) contains two free parameters $m$ and $\mu$. The most general interference of such solutions is obtained by summation over $m$ and integration over $\mu$ with a complex weight $g(m,\mu)$. Such an interference of more than one solutions can be utilized to engineer pulses with specific characteristics.
Specifically, in the left column of Fig.~\ref{fig:3} we interfere two solutions $\Psi_m$ with $m=0$ and $m=1$. As a result, the total power of the solution takes the form of an X-wave that respects parity. Note that the two solutions have a $\pi$-phase difference and thus exactly at $\tau=\zeta=0$ the amplitude is zero due to destructive interference. In the case of zero phase difference the interference of the two pulses is going to be constructive resulting to an X-wave that attains a maximum at $\tau=\zeta=0$. In the right column of Fig.~\ref{fig:3} we can see the X-wave generated by the interference of two solutions $A_m(\mu,\tau,\zeta)$ of the same order $m=1$ and different values of $\mu$. We observe destructive interference along and close to the center of the main lobes (having the form of notches) of the two pulses generating the X-wave.

\section{Conclusions}
In conclusion, we have found exact solutions that describe bidirectional pulses that propagate inside fiber Bragg gratings. They are obtained by directly solving the coupled-mode theory equations, describe bidirectional spatiotemporal X-waves or unidirectional pulses, and are expressed in terms of modified Bessel functions. We analyze the symmetries of the solutions, derive simplified asymptotic expressions and analyze the effect of the two free parameters in the form of the solutions. We also study waves that are obtained by the interference of such exact solutions.

\section{Acknowledgements}
N.K.E. is supported by the Erasmus Mundus NANOPHI Project (2013- 5659/002-001). 
N.S.N. is supported by the Onassis Public Benefit Foundation and by the Foundation for Education and European Culture (IPEP).

\newcommand{\noopsort[1]}{} \newcommand{\singleletter}[1]{#1}


\begin{thebibliography}{18}%
\makeatletter
\providecommand \@ifxundefined [1]{%
 \@ifx{#1\undefined}
}%
\providecommand \@ifnum [1]{%
 \ifnum #1\expandafter \@firstoftwo
 \else \expandafter \@secondoftwo
 \fi
}%
\providecommand \@ifx [1]{%
 \ifx #1\expandafter \@firstoftwo
 \else \expandafter \@secondoftwo
 \fi
}%
\providecommand \natexlab [1]{#1}%
\providecommand \enquote  [1]{``#1''}%
\providecommand \bibnamefont  [1]{#1}%
\providecommand \bibfnamefont [1]{#1}%
\providecommand \citenamefont [1]{#1}%
\providecommand \href@noop [0]{\@secondoftwo}%
\providecommand \href [0]{\begingroup \@sanitize@url \@href}%
\providecommand \@href[1]{\@@startlink{#1}\@@href}%
\providecommand \@@href[1]{\endgroup#1\@@endlink}%
\providecommand \@sanitize@url [0]{\catcode `\\12\catcode `\$12\catcode
  `\&12\catcode `\#12\catcode `\^12\catcode `\_12\catcode `\%12\relax}%
\providecommand \@@startlink[1]{}%
\providecommand \@@endlink[0]{}%
\providecommand \url  [0]{\begingroup\@sanitize@url \@url }%
\providecommand \@url [1]{\endgroup\@href {#1}{\urlprefix }}%
\providecommand \urlprefix  [0]{URL }%
\providecommand \Eprint [0]{\href }%
\providecommand \doibase [0]{http://dx.doi.org/}%
\providecommand \selectlanguage [0]{\@gobble}%
\providecommand \bibinfo  [0]{\@secondoftwo}%
\providecommand \bibfield  [0]{\@secondoftwo}%
\providecommand \translation [1]{[#1]}%
\providecommand \BibitemOpen [0]{}%
\providecommand \bibitemStop [0]{}%
\providecommand \bibitemNoStop [0]{.\EOS\space}%
\providecommand \EOS [0]{\spacefactor3000\relax}%
\providecommand \BibitemShut  [1]{\csname bibitem#1\endcsname}%
\let\auto@bib@innerbib\@empty
%</preamble>
\bibitem [{\citenamefont {Kashyap}(1999)}]{kashyap1999fiber}%
  \BibitemOpen
  \bibfield  {author} {\bibinfo {author} {\bibfnamefont {R.}~\bibnamefont
  {Kashyap}},\ }\href@noop {} {\emph {\bibinfo {title} {Fiber bragg
  gratings}}}\ (\bibinfo  {publisher} {Academic press},\ \bibinfo {year}
  {1999})\BibitemShut {NoStop}%
\bibitem [{\citenamefont {Agrawal}(2008)}]{agrawal-applications}%
  \BibitemOpen
  \bibfield  {author} {\bibinfo {author} {\bibfnamefont {G.~P.}\ \bibnamefont
  {Agrawal}},\ }\href {\doibase
  http://dx.doi.org/10.1016/B978-012374302-2.50001-2} {\emph {\bibinfo {title}
  {Applications of Nonlinear Fiber Optics}}},\ \bibinfo {edition} {2nd}\ ed.\
  (\bibinfo  {publisher} {Academic Press},\ \bibinfo {address} {Burlington},\
  \bibinfo {year} {2008})\BibitemShut {NoStop}%
\bibitem [{\citenamefont {Kogelnik}\ and\ \citenamefont
  {Shank}(1972)}]{kogel-jap1972}%
  \BibitemOpen
  \bibfield  {author} {\bibinfo {author} {\bibfnamefont {H.}~\bibnamefont
  {Kogelnik}}\ and\ \bibinfo {author} {\bibfnamefont {C.~V.}\ \bibnamefont
  {Shank}},\ }\href {\doibase 10.1063/1.1661499} {\bibfield  {journal}
  {\bibinfo  {journal} {Journal of Applied Physics}\ }\textbf {\bibinfo
  {volume} {43}},\ \bibinfo {pages} {2327} (\bibinfo {year} {1972})},\ \Eprint
  {http://arxiv.org/abs/http://dx.doi.org/10.1063/1.1661499}
  {http://dx.doi.org/10.1063/1.1661499} \BibitemShut {NoStop}%
\bibitem [{\citenamefont {Haus}\ and\ \citenamefont
  {Shank}(1976)}]{haus-ieeejqe1976}%
  \BibitemOpen
  \bibfield  {author} {\bibinfo {author} {\bibfnamefont {H.}~\bibnamefont
  {Haus}}\ and\ \bibinfo {author} {\bibfnamefont {C.}~\bibnamefont {Shank}},\
  }\href {\doibase 10.1109/JQE.1976.1069214} {\bibfield  {journal} {\bibinfo
  {journal} {IEEE Journal of Quantum Electronics}\ }\textbf {\bibinfo {volume}
  {12}},\ \bibinfo {pages} {532} (\bibinfo {year} {1976})}\BibitemShut
  {NoStop}%
\bibitem [{\citenamefont {Utaka}\ \emph {et~al.}(1984)\citenamefont {Utaka},
  \citenamefont {Akiba}, \citenamefont {Sakai},\ and\ \citenamefont
  {Matsushima}}]{utaka-el1984}%
  \BibitemOpen
  \bibfield  {author} {\bibinfo {author} {\bibfnamefont {K.}~\bibnamefont
  {Utaka}}, \bibinfo {author} {\bibfnamefont {S.}~\bibnamefont {Akiba}},
  \bibinfo {author} {\bibfnamefont {K.}~\bibnamefont {Sakai}}, \ and\ \bibinfo
  {author} {\bibfnamefont {Y.}~\bibnamefont {Matsushima}},\ }\href {\doibase
  10.1049/el:19840686} {\bibfield  {journal} {\bibinfo  {journal} {Electronics
  Letters}\ }\textbf {\bibinfo {volume} {20}},\ \bibinfo {pages} {1008}
  (\bibinfo {year} {1984})}\BibitemShut {NoStop}%
\bibitem [{\citenamefont {Legoubin}\ \emph {et~al.}(1995)\citenamefont
  {Legoubin}, \citenamefont {Boj}, \citenamefont {Delevaque}, \citenamefont
  {Douay}, \citenamefont {Bernage},\ and\ \citenamefont
  {Niay}}]{legoub-josaa1995}%
  \BibitemOpen
  \bibfield  {author} {\bibinfo {author} {\bibfnamefont {S.}~\bibnamefont
  {Legoubin}}, \bibinfo {author} {\bibfnamefont {S.}~\bibnamefont {Boj}},
  \bibinfo {author} {\bibfnamefont {E.}~\bibnamefont {Delevaque}}, \bibinfo
  {author} {\bibfnamefont {M.}~\bibnamefont {Douay}}, \bibinfo {author}
  {\bibfnamefont {P.}~\bibnamefont {Bernage}}, \ and\ \bibinfo {author}
  {\bibfnamefont {P.}~\bibnamefont {Niay}},\ }\href {\doibase
  10.1364/JOSAA.12.001687} {\bibfield  {journal} {\bibinfo  {journal} {J. Opt.
  Soc. Am. A}\ }\textbf {\bibinfo {volume} {12}},\ \bibinfo {pages} {1687}
  (\bibinfo {year} {1995})}\BibitemShut {NoStop}%
\bibitem [{\citenamefont {Hill}\ \emph {et~al.}(1987)\citenamefont {Hill},
  \citenamefont {Johnson}, \citenamefont {Bilodeau},\ and\ \citenamefont
  {Faucher}}]{hill-el1987}%
  \BibitemOpen
  \bibfield  {author} {\bibinfo {author} {\bibfnamefont {K.~O.}\ \bibnamefont
  {Hill}}, \bibinfo {author} {\bibfnamefont {D.~C.}\ \bibnamefont {Johnson}},
  \bibinfo {author} {\bibfnamefont {F.}~\bibnamefont {Bilodeau}}, \ and\
  \bibinfo {author} {\bibfnamefont {S.}~\bibnamefont {Faucher}},\ }\href
  {\doibase 10.1049/el:19870335} {\bibfield  {journal} {\bibinfo  {journal}
  {Electronics Letters}\ }\textbf {\bibinfo {volume} {23}},\ \bibinfo {pages}
  {465} (\bibinfo {year} {1987})}\BibitemShut {NoStop}%
\bibitem [{\citenamefont {Christodoulides}\ and\ \citenamefont
  {Joseph}(1989)}]{chris-prl1989}%
  \BibitemOpen
  \bibfield  {author} {\bibinfo {author} {\bibfnamefont {D.~N.}\ \bibnamefont
  {Christodoulides}}\ and\ \bibinfo {author} {\bibfnamefont {R.~I.}\
  \bibnamefont {Joseph}},\ }\href@noop {} {\bibfield  {journal} {\bibinfo
  {journal} {Phys. Rev. Lett.}\ }\textbf {\bibinfo {volume} {62}},\ \bibinfo
  {pages} {1746} (\bibinfo {year} {1989})}\BibitemShut {NoStop}%
\bibitem [{\citenamefont {Aceves}\ and\ \citenamefont
  {Wabnitz}(1989)}]{aceve-pla1989}%
  \BibitemOpen
  \bibfield  {author} {\bibinfo {author} {\bibfnamefont {A.~B.}\ \bibnamefont
  {Aceves}}\ and\ \bibinfo {author} {\bibfnamefont {S.}~\bibnamefont
  {Wabnitz}},\ }\href@noop {} {\bibfield  {journal} {\bibinfo  {journal} {Phys.
  Lett. A}\ }\textbf {\bibinfo {volume} {141}},\ \bibinfo {pages} {37}
  (\bibinfo {year} {1989})}\BibitemShut {NoStop}%
\bibitem [{\citenamefont {Winful}(1985)}]{winfu-apl1985}%
  \BibitemOpen
  \bibfield  {author} {\bibinfo {author} {\bibfnamefont {H.~G.}\ \bibnamefont
  {Winful}},\ }\href {\doibase 10.1063/1.95580} {\bibfield  {journal} {\bibinfo
   {journal} {Applied Physics Letters}\ }\textbf {\bibinfo {volume} {46}},\
  \bibinfo {pages} {527} (\bibinfo {year} {1985})},\ \Eprint
  {http://arxiv.org/abs/http://dx.doi.org/10.1063/1.95580}
  {http://dx.doi.org/10.1063/1.95580} \BibitemShut {NoStop}%
\bibitem [{\citenamefont {Lu}\ and\ \citenamefont
  {Greenleaf}(1992{\natexlab{a}})}]{lu-tuf1992a}%
  \BibitemOpen
  \bibfield  {author} {\bibinfo {author} {\bibfnamefont {J.-y.}\ \bibnamefont
  {Lu}}\ and\ \bibinfo {author} {\bibfnamefont {J.~F.}\ \bibnamefont
  {Greenleaf}},\ }\href@noop {} {\bibfield  {journal} {\bibinfo  {journal}
  {IEEE Trans. Ultrason. Ferroelectr. Freq. Control}\ }\textbf {\bibinfo
  {volume} {39}},\ \bibinfo {pages} {19} (\bibinfo {year}
  {1992}{\natexlab{a}})}\BibitemShut {NoStop}%
\bibitem [{\citenamefont {Lu}\ and\ \citenamefont
  {Greenleaf}(1992{\natexlab{b}})}]{lu-tuf1992b}%
  \BibitemOpen
  \bibfield  {author} {\bibinfo {author} {\bibfnamefont {J.-y.}\ \bibnamefont
  {Lu}}\ and\ \bibinfo {author} {\bibfnamefont {J.~F.}\ \bibnamefont
  {Greenleaf}},\ }\href@noop {} {\bibfield  {journal} {\bibinfo  {journal}
  {IEEE Trans. Ultrason. Ferroelectr. Freq. Control}\ }\textbf {\bibinfo
  {volume} {39}},\ \bibinfo {pages} {441} (\bibinfo {year}
  {1992}{\natexlab{b}})}\BibitemShut {NoStop}%
\bibitem [{\citenamefont {Saari}\ and\ \citenamefont
  {Reivelt}(1997)}]{saari-prl1997}%
  \BibitemOpen
  \bibfield  {author} {\bibinfo {author} {\bibfnamefont {P.}~\bibnamefont
  {Saari}}\ and\ \bibinfo {author} {\bibfnamefont {K.}~\bibnamefont
  {Reivelt}},\ }\href {\doibase 10.1103/PhysRevLett.79.4135} {\bibfield
  {journal} {\bibinfo  {journal} {Phys. Rev. Lett.}\ }\textbf {\bibinfo
  {volume} {79}},\ \bibinfo {pages} {4135} (\bibinfo {year}
  {1997})}\BibitemShut {NoStop}%
\bibitem [{\citenamefont {Porras}\ \emph {et~al.}(2003)\citenamefont {Porras},
  \citenamefont {Trillo}, \citenamefont {Conti},\ and\ \citenamefont
  {Trapani}}]{porra-ol2003}%
  \BibitemOpen
  \bibfield  {author} {\bibinfo {author} {\bibfnamefont {M.~A.}\ \bibnamefont
  {Porras}}, \bibinfo {author} {\bibfnamefont {S.}~\bibnamefont {Trillo}},
  \bibinfo {author} {\bibfnamefont {C.}~\bibnamefont {Conti}}, \ and\ \bibinfo
  {author} {\bibfnamefont {P.~D.}\ \bibnamefont {Trapani}},\ }\href {\doibase
  10.1364/OL.28.001090} {\bibfield  {journal} {\bibinfo  {journal} {Opt.
  Lett.}\ }\textbf {\bibinfo {volume} {28}},\ \bibinfo {pages} {1090} (\bibinfo
  {year} {2003})}\BibitemShut {NoStop}%
\bibitem [{\citenamefont {Conti}\ \emph {et~al.}(2003)\citenamefont {Conti},
  \citenamefont {Trillo}, \citenamefont {Di~Trapani}, \citenamefont {Valiulis},
  \citenamefont {Piskarskas}, \citenamefont {Jedrkiewicz},\ and\ \citenamefont
  {Trull}}]{conti-prl2003x}%
  \BibitemOpen
  \bibfield  {author} {\bibinfo {author} {\bibfnamefont {C.}~\bibnamefont
  {Conti}}, \bibinfo {author} {\bibfnamefont {S.}~\bibnamefont {Trillo}},
  \bibinfo {author} {\bibfnamefont {P.}~\bibnamefont {Di~Trapani}}, \bibinfo
  {author} {\bibfnamefont {G.}~\bibnamefont {Valiulis}}, \bibinfo {author}
  {\bibfnamefont {A.}~\bibnamefont {Piskarskas}}, \bibinfo {author}
  {\bibfnamefont {O.}~\bibnamefont {Jedrkiewicz}}, \ and\ \bibinfo {author}
  {\bibfnamefont {J.}~\bibnamefont {Trull}},\ }\href@noop {} {\bibfield
  {journal} {\bibinfo  {journal} {Phys. Rev. Lett.}\ }\textbf {\bibinfo
  {volume} {90}},\ \bibinfo {pages} {170406} (\bibinfo {year}
  {2003})}\BibitemShut {NoStop}%
\bibitem [{\citenamefont {Christodoulides}\ \emph {et~al.}(2004)\citenamefont
  {Christodoulides}, \citenamefont {Efremidis}, \citenamefont {Trapani},\ and\
  \citenamefont {Malomed}}]{chris-ol2004}%
  \BibitemOpen
  \bibfield  {author} {\bibinfo {author} {\bibfnamefont {D.~N.}\ \bibnamefont
  {Christodoulides}}, \bibinfo {author} {\bibfnamefont {N.~K.}\ \bibnamefont
  {Efremidis}}, \bibinfo {author} {\bibfnamefont {P.~D.}\ \bibnamefont
  {Trapani}}, \ and\ \bibinfo {author} {\bibfnamefont {B.~A.}\ \bibnamefont
  {Malomed}},\ }\href {http://ol.osa.org/abstract.cfm?URI=ol-29-13-1446}
  {\bibfield  {journal} {\bibinfo  {journal} {Opt. Lett.}\ }\textbf {\bibinfo
  {volume} {29}},\ \bibinfo {pages} {1446} (\bibinfo {year}
  {2004})}\BibitemShut {NoStop}%
\bibitem [{\citenamefont {Efremidis}\ \emph {et~al.}(2009)\citenamefont
  {Efremidis}, \citenamefont {Siviloglou},\ and\ \citenamefont
  {Christodoulides}}]{efrem-pla2009}%
  \BibitemOpen
  \bibfield  {author} {\bibinfo {author} {\bibfnamefont {N.~K.}\ \bibnamefont
  {Efremidis}}, \bibinfo {author} {\bibfnamefont {G.~A.}\ \bibnamefont
  {Siviloglou}}, \ and\ \bibinfo {author} {\bibfnamefont {D.~N.}\ \bibnamefont
  {Christodoulides}},\ }\href {\doibase DOI: 10.1016/j.physleta.2009.09.008}
  {\bibfield  {journal} {\bibinfo  {journal} {Phys. Lett. A}\ }\textbf
  {\bibinfo {volume} {373}},\ \bibinfo {pages} {4073 } (\bibinfo {year}
  {2009})}\BibitemShut {NoStop}%
\bibitem [{\citenamefont {Baronio}\ \emph {et~al.}(2016)\citenamefont
  {Baronio}, \citenamefont {Chen}, \citenamefont {Onorato}, \citenamefont
  {Trillo}, \citenamefont {Wabnitz},\ and\ \citenamefont
  {Kodama}}]{baron-ol2016}%
  \BibitemOpen
  \bibfield  {author} {\bibinfo {author} {\bibfnamefont {F.}~\bibnamefont
  {Baronio}}, \bibinfo {author} {\bibfnamefont {S.}~\bibnamefont {Chen}},
  \bibinfo {author} {\bibfnamefont {M.}~\bibnamefont {Onorato}}, \bibinfo
  {author} {\bibfnamefont {S.}~\bibnamefont {Trillo}}, \bibinfo {author}
  {\bibfnamefont {S.}~\bibnamefont {Wabnitz}}, \ and\ \bibinfo {author}
  {\bibfnamefont {Y.}~\bibnamefont {Kodama}},\ }\href {\doibase
  10.1364/OL.41.005571} {\bibfield  {journal} {\bibinfo  {journal} {Opt.
  Lett.}\ }\textbf {\bibinfo {volume} {41}},\ \bibinfo {pages} {5571} (\bibinfo
  {year} {2016})}\BibitemShut {NoStop}%
\end{thebibliography}
\end{document}